\documentclass[structabstract]{aa}  
\usepackage{graphicx}
\usepackage{txfonts}
\usepackage{natbib}       % for citep/citet

\def\degs{\ifmmode ^{\circ}\else$^{\circ}$\fi}
\def\amin{\ifmmode ^{\prime}\else$^{\prime}$\fi}
\def\asec{\ifmmode ^{\prime\prime}\else$^{\prime\prime}$\fi}
\newbox\grsign \setbox\grsign=\hbox{$>$}
\newdimen\grdimen \grdimen=\ht\grsign
\newbox\laxbox \newbox\gaxbox
\setbox\gaxbox=\hbox{\raise.5ex\hbox{$>$}\llap
     {\lower.5ex\hbox{$\sim$}}}\ht1=\grdimen\dp1=0pt
\setbox\laxbox=\hbox{\raise.5ex\hbox{$<$}\llap
     {\lower.5ex\hbox{$\sim$}}}\ht2=\grdimen\dp2=0pt
\def\gax{$\mathrel{\copy\gaxbox}$}
\def\lax{$\mathrel{\copy\laxbox}$}
\begin{document}
   \title{The nature of ``dark'' gamma-ray bursts}

   \author{J. Greiner\inst{1} \and
           T. Kr\"uhler\inst{1,2} \and
           S. Klose\inst{3} \and
%           others...
           P. Afonso\inst{1} \and
           C. Clemens\inst{1} \and
           R. Filgas\inst{1} \and
           D.H. Hartmann\inst{4} \and
           A. K\"{u}pc\"{u} Yolda\c{s}\inst{5} \and
           M. Nardini\inst{1} \and
           F.\,\,Olivares\,E.\inst{1} \and
           A. Rau\inst{1} \and
           A. Rossi\inst{3} \and
           P. Schady\inst{1} \and
           A. Updike\inst{4}
          }
   \institute{Max-Planck-Institut f\"ur extraterrestrische Physik,
              Giessenbachstrasse 1, 85748 Garching, Germany\\
         \email{jcg,kruehler,pafonso,cclemens,filgas,nardini,felipe,arau,pschady@mpe.mpg.de}
         \and
            Universe Cluster, Technische Universit\"{a}t M\"{u}nchen,
           Boltzmannstra{\ss}e 2, D-85748, Garching, Germany
         \and
             Th\"uringer Landessternwarte Tautenburg, Sternwarte 5,
             D-07778 Tautenburg,  Germany \\
             \email{klose,rossi@tls-tautenburg.de}
         \and
            Clemson Univ., Dept. of Physics and Astronomy, Clemson, SC 29634, 
             USA \\
             \email{dieter,aupdike@clemson.edu}
         \and
%             ESO, Karl-Schwarzschild-Str. 2, 85740 Garching, Germany \\
              Institute of Astronomy, University of Cambridge,
              Madingley Road, Cambridge CB3 0HA, U.K. \\
             \email{aky@ast.cam.ac.uk}
             }

   \date{Received 23 Jul 2010; in revised form 7 Oct. 2010, accepted 18 Oct 2010}

% \abstract{}{}{}{}{} 
% 5 {} token are mandatory

  \abstract
  % context heading (optional)
   {Thirteen years after the discovery of the first afterglows,
   the nature of dark gamma-ray bursts (GRB) still eludes explanation: 
   while each long-duration GRB typically has an X-ray afterglow, 
   optical/NIR emission is only seen for 40-60\% of them.
    }
  % aims heading (mandatory)
    {Here we use the afterglow detection statistics of the systematic follow-up
     observations performed with GROND since mid-2007 in order to derive the 
     fraction of ``dark bursts'' according to different methods,
     and to distinguish between various scenarios for ``dark bursts''.
    }
  % methods heading (mandatory)
   {Observations were performed with the 7-channel 
    ``Gamma-Ray Optical and Near-infrared Detector'' (GROND) at the 
    2.2m MPI/ESO telescope. We used the afterglow detection rate in dependence
    on the delay time between GRB and the first GROND exposure.
   }
  % results heading (mandatory)
    {For long-duration Swift bursts with a detected X-ray afterglow,
     we achieve a 90\%  (35/39)  detection rate of
     optical/NIR afterglows whenever our observations started within less than
     240\,min after the burst. Complementing our GROND data with Swift/XRT
     spectra we construct broad-band spectral energy distributions and
     derive rest-frame extinctions.
    }
    {We detect 25-40\% ``dark bursts'', depending on the definition used.
     The faint optical afterglow emission of ``dark bursts'' is mainly due to 
    a  combination of two contributing factors: 
    (i) moderate intrinsic extinction at moderate 
    redshifts, and (ii) about 22\% of ``dark'' bursts at redshift $>$5.
    }


   \maketitle
%
%________________________________________________________________

\section{Introduction}

Long-duration  gamma-ray   bursts  (GRBs)  are  the  high-energy
signatures of the  death of some massive stars, and they emit the bulk of
their radiation in the $\approx$300--800\,keV range.
For understanding the physics of the GRB explosion, the impact of
GRBs on their surrounding, as well as
the implications of GRBs on early star formation and cosmology, it is
crucial to observe their afterglow emission. While X-ray afterglows
are detected basically for each burst, the low detection rate of
optical/NIR afterglows has been a long standing problem in the GRB field
\citep{pac98, ggp98, ksm00, dfk01, fyn01, lcg02, pfg02, khg03, lfr06, jrw08, tlr08}, 
and it was first discussed systematically in \cite{fyn01} and \cite{lcg02}.

The  {\it Swift} satellite \citep{gcg04} was designed to slew
to GRB locations rapidly and provide positions with arcsec-accuracy through
observations of the afterglows with the X-ray telescope \citep[XRT,][]{bur05} 
and the UV-optical telescope \citep[UVOT,][]{rom05}. 
While the detection  of the  X-ray afterglows has dramatically facilitated 
the discovery of many new phenomena and increased the efficiency of ground-based
follow-up observations, the UVOT detection rate of afterglows is,
originally somewhat surprisingly, only $\sim $40\% \citep{rom06}.

The first burst in the \emph{afterglow era} for which no optical (or NIR)
afterglow was found
was GRB 970828 \citep{ggp98}.
Originally, those GRBs with X-ray but without optical afterglows
had been coined ``dark GRBs'' \citep{fyn01}. Later, this nomenclature was
made more specific by adding a time and brightness limit, e.g.
fainter than $R \sim$ 23 mag within 12 hrs of the burst.
As the next step, the basic prediction of the fireball scenario 
\citep{mr97} was utilized  and the optical-to-X-ray spectral index 
$\beta_{OX}$ (slope between the fluxes in the $R$-band and at 1 keV)
was used to define dark 
bursts \citep[$\beta_{OX} < 0.5$;][]{jak04}, or the
X-ray spectral and decay indices were used alternatively
to extrapolate the X-ray flux to the optical regime \citep{rol05}.
Ultimately, by using the X-ray flux and spectral ($\beta_{X}$) information 
from {\it Swift}, \cite{hkg09} propose to define dark bursts as those for which
$\beta_{OX}$ is shallower than ($\beta_{X} - 0.5$).

The darkness in the optical can
have several origins \citep[e.g.][]{fyn01}. The afterglow could
(i) have an intrinsically low luminosity, e.g. from
  an optically bright vs. optically dark dichotomy;
(ii) be strongly absorbed by intervening material, either very
  locally around the GRB or along the line-of-sight through the host galaxy,
or (iii) be at high redshift ($z>6$) so that Ly$\alpha$
  blanketing and absorption by the intergalactic medium
  would prohibit detection in the frequently used $R$ band \citep{lar00}.

Before the \textit{Swift} era, an analysis of a sub-sample of GRBs, 
namely those with particularly
accurate positions provided with the Soft X-ray Camera on HETE-2,
revealed optical afterglows for 10 out of 11 GRBs
\citep{villa04}. This suggested that the rapid availability of 
precise positions plays a major role in the identification of afterglows,
and that the majority of dark GRBs are neither at high redshift
nor strongly absorbed, 
but just dim, and/or rapidly decaying.
%i.e. the spread in afterglow
%brightness at a given time after the GRB is much larger than
%previous observations indicated.
However, {\it Swift} observations have provided 500 GRB localizations
at the few arcsec level within minutes of the trigger, and the fraction of 
non-detected afterglows is still about $\sim$30--40\%: UVOT detects about
40\% of the afterglows \citep{rko09}, and ground-based observations recover
another 20--30\% (see e.g. http:/$\!$/www.mpe.mpg.de/$\sim$jcg/grbgen.html).

This would imply that the accuracy and timeliness of a GRB position
is not the dominating factor. On the other hand, given the larger
mean redshift of GRBs in the {\it Swift} era \citep{ber05, jak06},
one could also argue that the mean flux of the afterglows is lower, 
and thus the effect of the better and faster \textit{Swift} localizations
on the afterglow detection rate is over-compensated by fainter afterglows
with respect to the above HETE-2 sub-sample.

Here, we use the afterglow detection rate of GROND, a de\-dicated ground-based 
GRB follow-up instrument, to derive new constraints on the fraction
of dark bursts. We complement the GROND data with \textit{Swift}/XRT spectra 
to construct broad-band spectral
energy distributions and to derive the rest-frame extinction.
Throughout this paper, we use the definition F$_{\nu} \propto \nu^{-\beta}$.

\section{GROND observation statistics}

GROND, a simultaneous 7-channel optical/near-infrared ima\-ger \citep{gbc08} 
moun\-ted at  the 2.2\,m  MPI/ESO telescope at La Silla (Chile), started 
operation in May 2007. GROND has been built as a dedicated GRB follow-up 
instrument
and has observed basically every GRB visible from La Silla (weather allowing) 
since April 2008.

GROND observations of GRBs within the first day are fully automated
(see \citealt{gbc08} for more details). The fastest reaction on a 
{\it Swift} alert so far was 140 s (Swift trigger 353627); 
more typical times are 200--300 s for night-time bursts.
The two dominant contributors for this delay are the read-out time of the 
interrupted exposure (particularly the Wide-Field Imager, one of two other 
instruments at the telescope), and the dome rotation 
speed. The distribution of delay times of GRBs with GROND is given
in Tab. \ref{frac} (last column). Note that due to La Silla being located
at similar geographic longitude as the South Atlantic Anomaly (where
most gamma-ray detectors need to switch off their high-voltage),
the fraction of night-time bursts is nearly a factor two lower than
for observatories at other geographic longitudes.

It is obvious that the later after a GRB trigger the GROND observation starts, 
the lower is the fraction of afterglow detections
(Tab. \ref{frac}).
This is expected and readily explained by the rapid fading of 
GRB afterglows and the limiting sensitivity of the instrument/telescope
in use. What is surprising, however, is the high detection rate in the first 
two time bins -- this will be discussed in the following.
The relatively high detection rate at $>$24 hrs is
     biased by the Fermi/LAT bursts which are, on average, more energetic
     with correspondingly brighter afterglows \citep{mkr10}.

\begin{table}[th]
%   \vspace{-0.2cm}
   \caption{GROND afterglow detection fraction of long-duration bursts
     as a function of time delay of the start of the observation after the 
     GRB trigger. }
   \vspace{-0.2cm}
%   \begin{center}
      \begin{tabular}{ccc}
      \hline
      \noalign{\smallskip}
      Delay (hrs) & detected vs. observed (fraction) & fraction of total \\ 
      \noalign{\smallskip}
      \hline
      \noalign{\smallskip}
      $<$0.5 & 20 /22 (91\%) & 17\% \\
     0.5--4  & 15 /17 (88\%) & 14\% \\
       4--8  & 10 /21 (48\%) & 16\% \\
       8--16 & 22 /36 (61\%) & 28\% \\
      16--24 & 13 /22 (59\%) & 17\% \\
     $>$ 24  &  5 /10 (50\%) & ~8\% \\
      \noalign{\smallskip}
      \hline
   \end{tabular}
    \tablefoot{Based on a total of 128 bursts
      observed between 070802 and (inclusive) 100331.}
%   \end{center}
   \label{frac}
   %\vspace{-.35cm}
\end{table}

Table \ref{grb30min} contains all those bursts which have been observed with 
GROND within less than 240 min (4 hrs) after the {\it Swift} trigger and 
which have 
XRT-detected afterglows (until 31 March 2010). First we note that 
(within this sample) there is a clear difference between 
the detection 
likelihood for long- and short-duration bursts. While we detect only
1 out of 4 short bursts, it is 35 out of 39 for long-duration bursts.
This resembles the well-known fact that short-duration bursts have
generally much fainter afterglows than long-duration bursts.
Second, we note that a further down-selection of long-duration bursts at
high Galactic latitude does not change our detection rate.
This is due to GROND's capability of imaging in all bands simultaneously,
in particular also in the near-infrared.
Third, we emphasize that this sample lacks any selection effects 
beyond requiring an X-ray afterglow.

\section{Fitting broad-band spectral energy distributions}

\subsection{Overall properties}

GROND and \textit{Swift}/XRT data have been reduced in the standard manner 
using 
pyraf/IRAF \citep{tod93, kkg08b} for the optical/NIR data and the XRT pipeline 
for the X-ray data. The optical/NIR imaging was calibrated against the primary 
SDSS standard star network, or cataloged magnitudes of field stars from the 
SDSS in the case of $g^\prime r^\prime i^\prime z^\prime$ observations or the 
2MASS catalog for $JHK_S$ imaging. This results in typical absolute accuracies 
of $\pm$0.03~mag in $g^\prime r^\prime i^\prime z^\prime$ and $\pm$0.05~mag in 
$JHK_S$. X-ray data were cleaned for time intervals of flaring activity and 
the early steep decay. The X-ray spectra were flux normalized to the epoch 
corresponding to the GROND observations using the XRT light curves from 
\citet{ebp07, ebp09}. This common reference time was selected to be after 
the early optical rise observed in some light curves and is different for 
each burst, but typically during the first few hours after the \textit{Swift} 
trigger.

\begin{figure*}[ht]
  \vspace{-0.3cm}
\includegraphics[width=12.0cm]{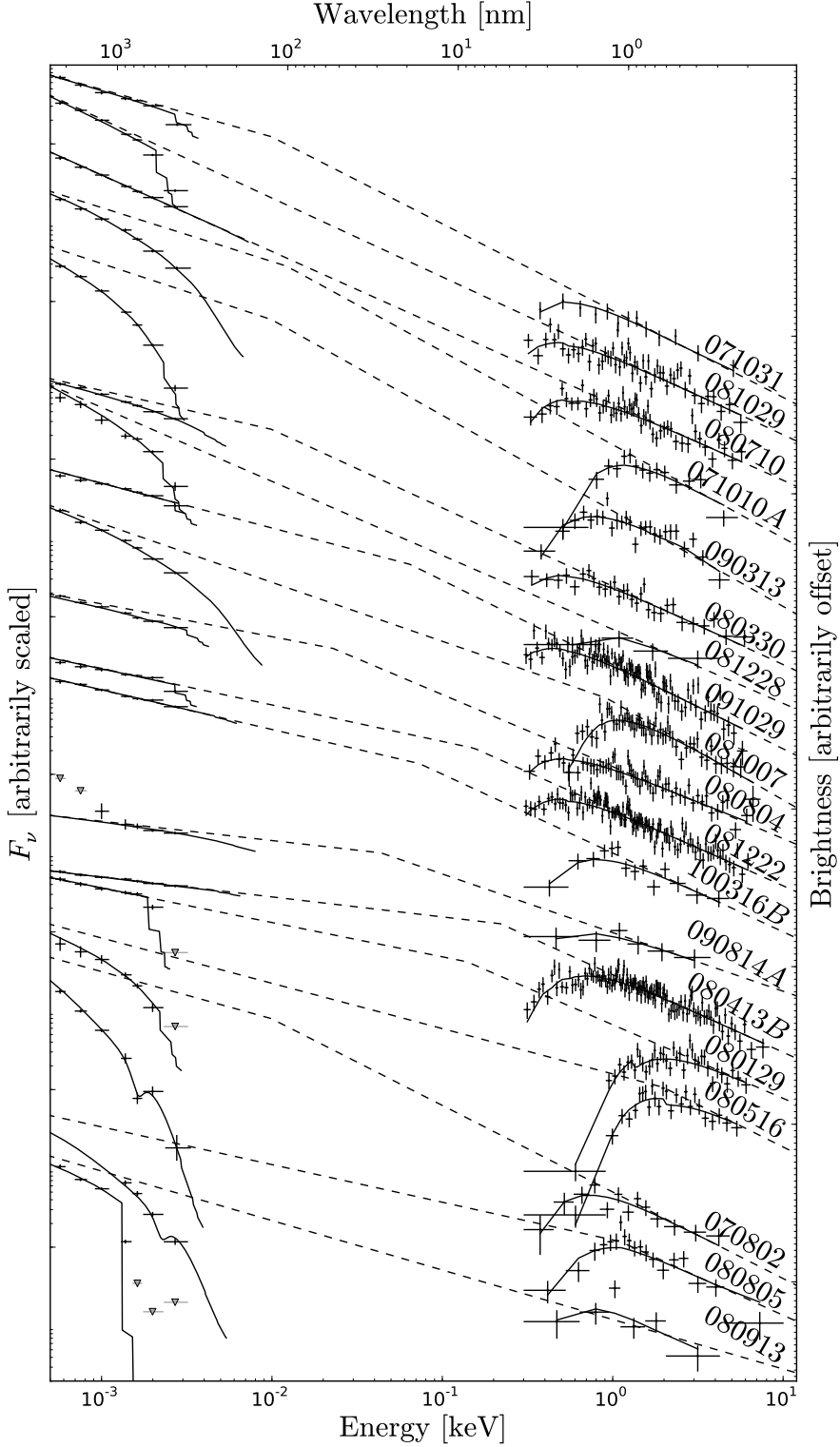}
\hfill\parbox[t]{5.7cm}{\vspace{-7.7cm}\caption[AllSED]{Spectral energy 
   distributions of a selection of the long-duration GRBs with redshift 
   in our sample, labeled on the 
   right side. The X-ray spectrum has been extracted from early
   times, but after the rapid fading and excluding flares, and then shifted
   in mid-time to the GROND mid-time (see text for details).
   The dashed lines indicate the best-fit model for each burst, with the 
   break between the X-ray and optical/NIR always treated as a free parameter.
   Curvature in the X-rays is due to Galactic plus host-intrinsic
   absorption, and curvature in the 
   observed optical/NIR range due to host-intrinsic extinction (data have
   been corrected for Galactic foreground extinction before fitting).
   For better visibility, the intensity scaling has been choosen to minimize 
   overlap of the different SEDs, and thus is completely arbitrary. 
  \label{allsed}}}
\end{figure*}

We then combined XRT and Galactic foreground extinction \citep{sfd98}
corrected GROND data to establish a broad-band spectral energy 
distribution (SED) -- see Fig. \ref{allsed}. 

\begin{table*}[th]
   \caption{GRBs observed with GROND within 240 min after
     the {\it Swift} trigger (including 39 long- and 4 short-duration bursts). 
      }
   \vspace{-0.3cm}
      \begin{tabular}{lccrrcclllccc}
     \hline
      \noalign{\smallskip}
 GRB &  D  & B~~ & Delay  & Dur &  AG  & Refs.$^{1)}$ &~~~$z^{2)}$ & Comment
    & $\beta_O$ & $\beta_X$ & $A_V$(host) & $N_H$(host) \\
     &   &   & (min)&(min)& (G/O)&        &              &   
    &           &           & (mag) & (10$^{21}$ cm$^{-2}$)\\    
      \noalign{\smallskip}
      \hline
      \noalign{\smallskip}
070802  &L& --57$\degr$ &  16.0 &  72 & y/y &  (1) & 2.45  (1)  &        & 0.60 & 1.10$^{+0.14}_{-0.12}$ & 1.23$^{+0.18}_{-0.16}$ & 20$^{+7}_{-8}$ \\
071010A &L& --18$\degr$ &  16.4 &  25 & y/y &  (2) & 0.98  (2)  &        & 0.68 & 1.18$^{+0.13}_{-0.09}$ & 0.45$^{+0.11}_{-0.18}$ & 17$^{+8}_{-5}$\\
071031  &L& --59$\degr$ &   3.7 & 465 & y/y &  (3) & 2.69  (3)  & UVOT   & 0.60 & 1.10$^{+0.11}_{-0.07}$ & 0.02$^{+0.03}_{-0.02}$ & 10$^{+9}_{-5}$\\
080129  &L& --1$\degr$  &   3.5 & 100 & y/y &  (4) & 4.35  (4)  & UVOT   & 0.42 & 0.92$^{+0.12}_{-0.13}$ & 0.00$^{+0.06}_{-0.00}$ & 240$^{+110}_{-120}$ \\
080210  &L& +34$\degr$  &  72.1 &  53 & y/y &  (5) & 2.64 (5,6) & UVOT   & 0.76 & 1.26$^{+0.03}_{-0.03}$ & 0.24$^{+0.03}_{-0.03}$ & 14$^{+11}_{-8}$ \\
080218B &L&  +9$\degr$  &  40.5 & 496 & n/n &  (6) & --          &       &  --- & 1.29$^{+0.45}_{-0.38}$ & --- & --- \\
080330  &L&  +69$\degr$ &   3.3 & 117 & y/y &  (7) & 1.51  (7)   & UVOT  & 0.49 & 0.99$^{+0.09}_{-0.11}$ & 0.10$^{+0.03}_{-0.08}$ & 3.1$^{+1.8}_{-1.4}$\\
080411  &L& --44$\degr$ & 122.7 &  94 & y/y &  (8) & 1.03 (8,6) & UVOT   & 0.42 & 0.92$^{+0.06}_{-0.06}$ & 0.00$^{+0.00}_{-0.02}$ & 5.8$^{+0.5}_{-0.5}$ \\
080413A &L& --16$\degr$ & 140.0 & 213 & y/y &  (9) & 2.43 (9,6) & UVOT   & 0.67 & 1.17$^{+0.09}_{-0.03}$ & 0.03$^{+0.03}_{-0.03}$ & 12$^{+11}_{-8}$ \\
080413B &L& --47$\degr$ &   5.6 &  98 & y/y & (10) & 1.10 (10,6) & UVOT & 0.25$^{+0.07}_{-0.18}$ & 0.97$^{+0.07}_{-0.07}$ & 0.00$^{+0.13}_{-0.00}$ & 4.1$^{+1.0}_{-1.0}$ \\
080516  &L&  +2$\degr$  &   8.5 & 102 & y/n & (11) & 3.6p (11) &        & 0.59 & 1.09$^{+0.08}_{-0.14}$ & 0.43$^{+0.08}_{-0.08}$ & 240$^{+120}_{-100}$ \\
080520  &L& -20$\degr$  & 180.6 & 190 & y/y & (12) & 1.54 (12,6) & UVOT & 0.57 & 1.07$^{+0.29}_{-0.10}$ & 0.53$^{+0.40}_{-0.42}$ & 17$^{+13}_{-6}$ \\
080605  &L& +20$\degr$  &  82.8 & 469 & y/y & (13) & 1.64 (13,6) & UVOT & ---  & 0.67$^{+0.01}_{-0.01}$ & 0.47$^{+0.03}_{-0.03}$ & 10.1$^{+0.9}_{-0.8}$ \\
080707  &L& --27$\degr$ &  92.0 &  83 & y/y & (14) & 1.23 (14,6) & UVOT & 0.64 & 1.14$^{+0.05}_{-0.18}$ & 0.11$^{+0.14}_{-0.08}$ & 5.6$^{+3.0}_{-3.0}$ \\
080710  &L& --43$\degr$ &   6.3 & 218 & y/y & (15) & 0.85 (15,6) & UVOT & ---  & 0.97$^{+0.02}_{-0.02}$ & 0.00$^{+0.02}_{-0.00}$  & 1.4$^{+0.6}_{-0.5}$ \\
080804  &L& --48$\degr$ &  18.8 &  67 & y/y & (16) & 2.20  (16,6) & UVOT & 0.43 & 0.93$^{+0.09}_{-0.07}$ & 0.06$^{+0.04}_{-0.04}$ & 4.8$^{+2.7}_{-2.0}$ \\
080805  &L& --38$\degr$ &   4.0 &   1 & y/y & (17) & 1.50  (17,6) &      & 0.47 & 0.97$^{+0.05}_{-0.05}$ & 1.01$^{+0.19}_{-0.14}$ & 10$^{+6}_{-4}$\\
080913  &L& --43$\degr$ &   4.5 & 227 & y/y & (18) & 6.70  (18) &        & ---  & 0.67$^{+0.05}_{-0.05}$ & 0.13$^{+0.06}_{-0.07}$  & 0$^{+130}_{-0}$\\
080915  &L& --41$\degr$ &   5.8 & 157 & n/n & (19) &  ---  & X-ray faint & ---  & 1.20$^{+2.81}_{-0.71}$ & ---             & ---\\
080919  &S& --06$\degr$ &   8.3 & 112 & n/n & ---  &  ---      &         & ---  & --- & --- & --- \\
081007  &L& --60$\degr$ &  14.5 & 137 & y/y & (20) & 0.53 (19) & UVOT    & 0.75 & 1.25$^{+0.03}_{-0.06}$ & 0.36$^{+0.06}_{-0.04}$ & 9.1$^{+1.5}_{-1.3}$\\
081008  &L& -21$\degr$  & 225.9 & 280 & y/y & (21) & 1.97 (20) & UVOT    & 0.56 & 1.06$^{+0.11}_{-0.04}$ & 0.08$^{+0.04}_{-0.08}$ & 5.4$^{+3.8}_{-2.6}$ \\
081029  &L& --46$\degr$ &   6.1 & 347 & y/y & (22) & 3.85 (21) & UVOT    & ---  & 1.00$^{+0.01}_{-0.01}$ & 0.03$^{+0.02}_{-0.03}$ & 7.9$^{+6.8}_{-5.9}$\\
081121  &L& --30$\degr$ & 233.9 & 511 & y/y & (23) & 2.51 (22) & UVOT    & 0.36 & 0.86$^{+0.02}_{-0.04}$ & 0.00$^{+0.03}_{-0.00}$ & 2.4$^{+2.0}_{-1.8}$ \\
081222  &L& --79$\degr$ &  14.9 &  26 & y/y & (24) & 2.77 (23) & UVOT    & 0.47 & 0.97$^{+0.05}_{-0.05}$ & 0.00$^{+0.03}_{-0.00}$ & 6.4$^{+2.9}_{-2.7}$\\
081226  &S& --19$\degr$ &  11.2 & 211 & n/n & (25) & ---       &         & ---  & --- & --- & --- \\
081228  &L& --27$\degr$ &   7.2 & 105 & y/n & (26) & 3.4p (24) &         & ---  & 0.93$^{+0.04}_{-0.04}$ & 0.12$^{+0.06}_{-0.08}$ & 13$^{+50}_{-13}$\\
090102  &L& +35$\degr$  & 150.2 &  50 & y/y & (27) & 1.55  (25) & UVOT   & 0.35 & 0.85$^{+0.10}_{-0.07}$ & 0.45$^{+0.06}_{-0.08}$ & 10$^{+3}_{-2}$ \\
090305  &S& +15$\degr$  &  27.9 & 102 & y/y & (28) &  ---  & X-ray faint  & ---  & --- & --- & --- \\
%090307 &L& +15$\degr$  &  80.0 &  66 & n/n & (?)  & --- (?)    & UVOT??  &  & 0.??$^{+0.??}_{-0.??}$ & 0.??$^{+0.??}_{-0.??}$ & ??$^{+??}_{-??}$ \\
090313  &L& +70$\degr$  &   7.0 &  78 & y/y & (29) & 3.38 (26) &         & 0.71 & 1.21$^{+0.09}_{-0.05}$ & 0.42$^{+0.06}_{-0.05}$ & 41$^{+12}_{-10}$\\
090429B &L& +74$\degr$  &  13.0 &  12 & n/y & (30) & 9.2p (27)  &        & ---  & 0.95$^{+0.25}_{-0.23}$ & --- & ---\\
090519  &L& +35$\degr$  & 104.2 & 101 & y/y & (31) & 3.85 (28)  & UVOT   & ---  & 0.71$^{+0.13}_{-0.04}$ & 0.01$^{+0.19}_{-0.01}$ & 55$^{+35}_{-21}$ \\
090812  &L& --65$\degr$ & 131.0 & 173 & y/y & (32) & 2.45 (29)  & UVOT   & 0.36 & 0.86$^{+0.12}_{-0.12}$ & 0.41$^{+0.04}_{-0.09}$ & 10$^{+7}_{-7}$ \\
090814  &L& +48$\degr$  &  14.1 &  54 & y/y & (33) & 0.70 (30) & UVOT    & 0.26 & 0.76$^{+0.27}_{-0.30}$ & 0.05$^{+0.14}_{-0.05}$ & 0.6$^{+4.1}_{-0.6}$\\
090904B$^{3)}$&L&+4$\degr$& 3.6 & 175 & y/n & (34) & $<$5.0p     &       & 0.46 & 0.96$^{+0.18}_{-0.25}$ & --- & ---\\
090926B &L& --60$\degr$ & 236.2 & 225 & y/y & (35) & 1.24  (31) &        & ---  & 0.73$^{+0.09}_{-0.07}$ & 1.42$^{+1.08}_{-0.57}$ & 22$^{+5}_{-4}$ \\
091018  &L& --57$\degr$ & 188.5 & 169 & y/y & (36) & 0.97  (32) & UVOT   & 0.54 & 1.04$^{+0.10}_{-0.06}$ & 0.08$^{+0.07}_{-0.08}$ & 2.8$^{+1.0}_{-0.7}$ \\
091029  &L& --46$\degr$ &   4.6 & 158 & y/y & (37) & 2.75 (33) & UVOT    & 0.57 & 1.07$^{+0.05}_{-0.05}$ & 0.00$^{+0.04}_{-0.00}$ & 6.5$^{+3.0}_{-2.5}$\\
091127  &L& --67$\degr$ &  58.3 & 316 & y/y & (38) & 0.49  (34) & UVOT   & 0.27 & 0.77$^{+0.03}_{-0.05}$ & 0.00$^{+0.04}_{-0.00}$ & 1.0$^{+0.4}_{-0.4}$ \\
091221  &L& --25$\degr$ & 236.1 & 129 & y/y & (39) & $<$3.3p   & UVOT    & ---  & 0.76$^{+0.05}_{-0.05}$ & --- & --- \\
100117A &S&--64$\degr$  & 220.1 &  70 & n/y & (40) & 0.92 (35) &         & ---  & --- & --- & --- \\
100205A &L& +45$\degr$  & 149.2 &  63 & n/y & (41) & --        &         & ---  & 1.08$^{+0.27}_{-0.32}$ & --- & --- \\
100316B &L& +13$\degr$  &  15.1 &  92 & y/y & (42) & 1.18 (36) & UVOT    & 0.50 & 1.00$^{+0.05}_{-0.09}$ & 0.00$^{+0.05}_{-0.00}$ & 2.9$^{+2.7}_{-2.1}$\\
      \noalign{\smallskip}
      \hline
      \noalign{\smallskip}
   \end{tabular}

   \tablefoot{Columns 2--10 are the duration (D) classification of 
     the GRB according to the canonical long (L) / short (S) scheme 
     \citep{kou93}, Galactic Latitude $B$, the delay between the GRB trigger 
     and the start of the GROND observation 
     (which is the sum of the delay between 
     the burst and the arrival of the {\it Swift} notification at the GROND 
     computer, and that until the GROND start), the duration (Dur) of the 
     GROND observation during the first night, the afterglow (AG) detection
     by GROND (G) or others (O), references for the afterglow observations,
     the redshift $z$ with  reference, and special comments (UVOT
     detection or X-ray brightness).
     The last four columns are the best fit spectral slopes in the
     optical/NIR ($\beta_O$) and X-ray ($\beta_X$) band, the rest-frame
     extinction $A_V$ and absorption $N_H$ from the combined 
     GROND/{\it Swift}-XRT spectral fit. $\beta_O$ was generally
     fixed to $\beta_X$ ($-$0.5) in the fit, and thus has the same error as
     $\beta_X$.}

    $^{1)}$ References for previously reported GROND afterglow detections
            as well as those from other groups for those GRBs which have
            not been detected by GROND:
            %(1) \cite{hak07}   (2) \cite{gck07}   
           (1) \cite{kkg08}, %\cite{bem07}
           (2) \cite{cdk08},        (3) \cite{kgm09},   (4) \cite{gkm09},
           (5) \cite{kgy08},        (6) \cite{rgky08}, 
           (7) \cite{ckg08, gck09}, (8) \cite{kkgcy08}, (9) \cite{ryr08},
          (10) \cite{kgk08},      (11) \cite{fkjg08},  (12) \cite{rfk08},
          (13) \cite{ckgk08},     (14) \cite{clgy08},  (15) \cite{kga10},  
          (16) \cite{ksg08},      (17) \cite{ksgc08}, (18) \cite{rgk08, gkr08},
          (19) \cite{rkg08},      (20) \cite{dbm08},  (21) \cite{ysr10},
          (22) \cite{clg08},      (23) \cite{lkg08}, 
          (24) \cite{uac08},  (25) \cite{akk08},  (26) \cite{akkg08},
          (27) \cite{akk09, gkp10}, (28) \cite{ccp09},
          (29) \cite{ukc09},  (30) \cite{okg09},    %(24) \cite{okk09}
          %(25) \cite{rog09}  
          (31) \cite{rkg09},  (32) \cite{urk09},
          (33) \cite{uoa09},  (34) \cite{oakg09},  (35) \cite{mfd09},
          (36) \cite{fky09},  (37) \cite{fug09},   (38) \cite{urr09},
          (39) \cite{fak09},  (40) \cite{lgf10},  (41) \cite{nkr10, tlc10},
          (42) \cite{afg10} 
           \\
    $^{2)}$ A ``p'' after the redshift value indicates a photometric redshift 
           estimate. References for the redshifts are: 
            %(1) \cite{pcb09}   
           (1) \cite{ptm07, efh09},  (2) \cite{ppm07},
           (3) \cite{ljj07, flv08},  (4) \cite{gkm09},  (5) \cite{jvm08},
           (6) \cite{fjp09},
           (7) \cite{mfj08, gck09},  (8) \cite{tcm08},  (9) \cite{tmv08}, 
          (10) \cite{vtm08},        (11) \cite{fkg08}, (12) \cite{jfm08},
          (13) \cite{jvx08},        (14) \cite{fmm08}, (15) \cite{pcb08},
          (16) \cite{tuv08},        (17) \cite{jfv08}, (18) \cite{gkf09},
          (19) \cite{bfc08},        (20) \cite{ddc08}, (21) \cite{dcd08}, 
          (22) \cite{ber08},        (23) \cite{cfc08}, (24) \cite{akg08, ksg10},
          (25) \cite{ujm09}, (26) \cite{cpc09, ugt10}, (27) \cite{clf10, tan10},
           %(19) \cite{okk09}
          (28) \cite{tjc09},        (29) \cite{ugf09}, (30) \cite{jug09},   
          (31) \cite{fmj09},        (32) \cite{chs09}, (33) \cite{cpco09}, 
          (34) \cite{cfl09},        (35) \cite{fong10}, (36) \cite{vdm10}.  \\
     $^{3)}$ The values for host $A_V$ and $N_H$ are given for redshift zero.
   \label{grb30min}
\end{table*}

\clearpage

These SEDs have been fit with two alternative models: 
(i) a single power law with free slope and normalization, plus free 
   source-intrinsic extinction  of SMC/LMC or MW-type (for the GROND data) 
  and Galactic plus rest-frame equivalent neutral hydrogen 
  column density (for the X-ray data) 
  assuming solar abundance; or 
(ii) a broken power law where the break energy is left free but
  the difference in the two slopes is fixed to 0.5,
  and all other parameters are left free as above. Note that fixing the break
does not imply that the optical/NIR SED is not fit; instead, it is fit
together with the X-ray SED (the X-ray slope in general does not dominate
the SED fit). These models provide a very 
good fit to all but one afterglow SED (Fig. \ref{allsed}): For this
exceptional case, GRB~080413B, 
the combined GROND/XRT SED requires a difference $>$0.5 between the low and 
high energy slopes of the synchrotron emission model \citep{fil10}. 
Except for 
seven GRBs (080605, 080710, 080913, 081029, 081228, 090926B, 091221)
all GRBs are better fit with a break between the X-ray and optical/NIR. 
The best-fit parameters are listed in the last four columns of 
Tab. \ref{grb30min}.
With the exception of 
five bursts (GRBs 070802, 080210, 080605, 080805, 090102), 
the SEDs of the afterglows are consistent with being reddened with an 
SMC extinction law.

\subsection{Notes on individual GRBs}
{\bf 070802} has a $\beta_{\rm OX}=0.5$, but since the best-fit location of the
  spectral break is just bluewards of the $g'$-band, it still requires
  a substantial $A_{\rm V} \sim 1.2$ mag. This burst is also an example of a
  border-line case: its $\beta_{\rm OX}$ just misses the ``dark burst''
  criterion of \cite{jak04}, but due to its $\beta_{\rm X}$ being slightly
  steeper than 1.0, it qualifies as ``dark burst'' according to 
  \cite{hkg09}. \\
{\bf 080218B:}  
  If we assume a break of 0.5 between the X-ray 
  and optical/NIR band, as for the majority of the bursts in our sample,
  the GROND upper limits allow the following
   $(z,A_V)$ pairs of e.g. 
  (3.5,1.5) or (5,1.0) or (7,0.7) or (10,0.5)
   as  explanations of the non-detection. \\
{\bf 080516:} With the improved calibration, our best-fit redshift is
  $z$=3.6$\pm$0.6. Since GROND is not sensitive to redshifts smaller
  than $\approx$3, the above redshift is consistent with zero at the
  90\% confidence level. However, we use this best-fit redshift to derive 
  $A_{\rm V}$ which therefore could be a lower limit. \\
{\bf 080805:} In this case there are no simultaneous GROND and XRT observations
  of the afterglow.
  The flux normalization of the XRT spectrum was obtained by back-extrapolating
  the XRT afterglow light-curve to the earlier time of GROND observations. \\
{\bf 080915}   has  a very faint X-ray afterglow \citep{oue08}. 
 If we assume a break of 0.5 between the X-ray 
  and optical/NIR band, as for the majority of the bursts in our sample,
  we predict $r' \sim 23.2$ mag with  $A_{\rm V}$ = 0.  This is brighter than 
 our GROND detection limit ($r' > 25.2$ mag) for this burst.
  Thus, various $(z,A_V)$ pairs of e.g. 
  (3.5,$>$0.5) or (5,$>$0.25) or (7,$>$0.1)
  provide good explanations of the non-detection (Fig. \ref{Avz}). \\
{\bf 090429B:} Due to bad sky conditions, this burst was not seen in the first
  night, but detected in the second night at the 2$\sigma$ level in the 
  $H$ band, but not in $J$. Since GROND is more sensitive in $J$ as compared
  to $H$, this is consistent with a break between $J$ and $H$, and thus
  with the photometric redshift determined by \cite{clf10}. \\
{\bf 090904B:} The object is detected in all GROND filters except the g'-band
  which implies a secure upper limit on the redshift of $z<5$. 
  However, the $g'$-band non-detection is also well explained by the large 
  Galactic foreground reddening ($E(B-V)$ = 1.76 mag \cite{sfd98}). 
  With no accurate redshift available, the best-fit extinction 
  of $A_{\rm V} = 2$ mag at $z=0$ is an upper limit.
  However, the large foreground extinction of $A_{\rm V} \sim 5$ mag can be
  expected to come with a large (systematic) error, which propagates also
  to our fit values; 
  thus we cannot distinguish between Galactic foreground and host
  extinction. \\
{\bf 090926B:} While GROND observations of this GRB started quickly, clouds
  prevented continued observations after a few minutes. Further observations
  were done starting 6.2 hrs after the burst, and additionally on 14 Feb 2010.
  The latter observation confirmed the fading of the candidate counterpart
  of \cite{mfd09}, and the association to the host galaxy as well as the
  redshift determination via host spectroscopy \citep{fmj09}. 
 \\
{\bf 091221:} No spectroscopic redshift is available. The afterglow is detected
  in all GROND filters, and the 3$\sigma$ upper limit on the redshift is 
  z$<$3.3. 
  For a putative redshift of 2, we obtain
  $A_{\rm V} = 0.21^{+0.13}_{-0.12}$ mag and  
  $N_{\rm H} = 1^{+14}_{-1}$ 10$^{21}$ cm$^{-2}$.
 \\
{\bf 100117A:} The redshift of $z$=0.92 is contained in a summary table in
  \cite{ber10} and referenced there as \citet[][in prep.]{fong10}. \\
{\bf 100205A:} Due to cloud coverage, GROND observations started only
   2.3\,hr after the burst, and then were severely affected by passing cirrus.
   Correspondingly, no detection was achieved, and the upper limits obtained 
   by GROND were about 2 magnitudes worse than the nominal sensitivity
   under normal conditions \citep{nkr10}. 
   At about the same time (2.6\,hr after the GRB),
   Gemini observations revealed a near-infrared counterpart \cite{tlc10} with
   a very red colour of $H - K = 1.6 \pm 0.5$ mag (AB system) \citep{cft10}.
   If this colour is due to Lyman-$\alpha$ absorption within the $H$ filter 
   bandpass, then this would imply a redshift of 11 \lax\ $z$ \lax\ 13.5. 
   Lower-redshift solutions with significant local (host) extinction, 
   unusual afterglow colours, or substantial contributions from an 
   underlying host galaxy have not been excluded though \citep{cft10}.

\section{Discussion}

\subsection{The nature of dark bursts}

\begin{figure}[ht]
\includegraphics[width=\columnwidth]{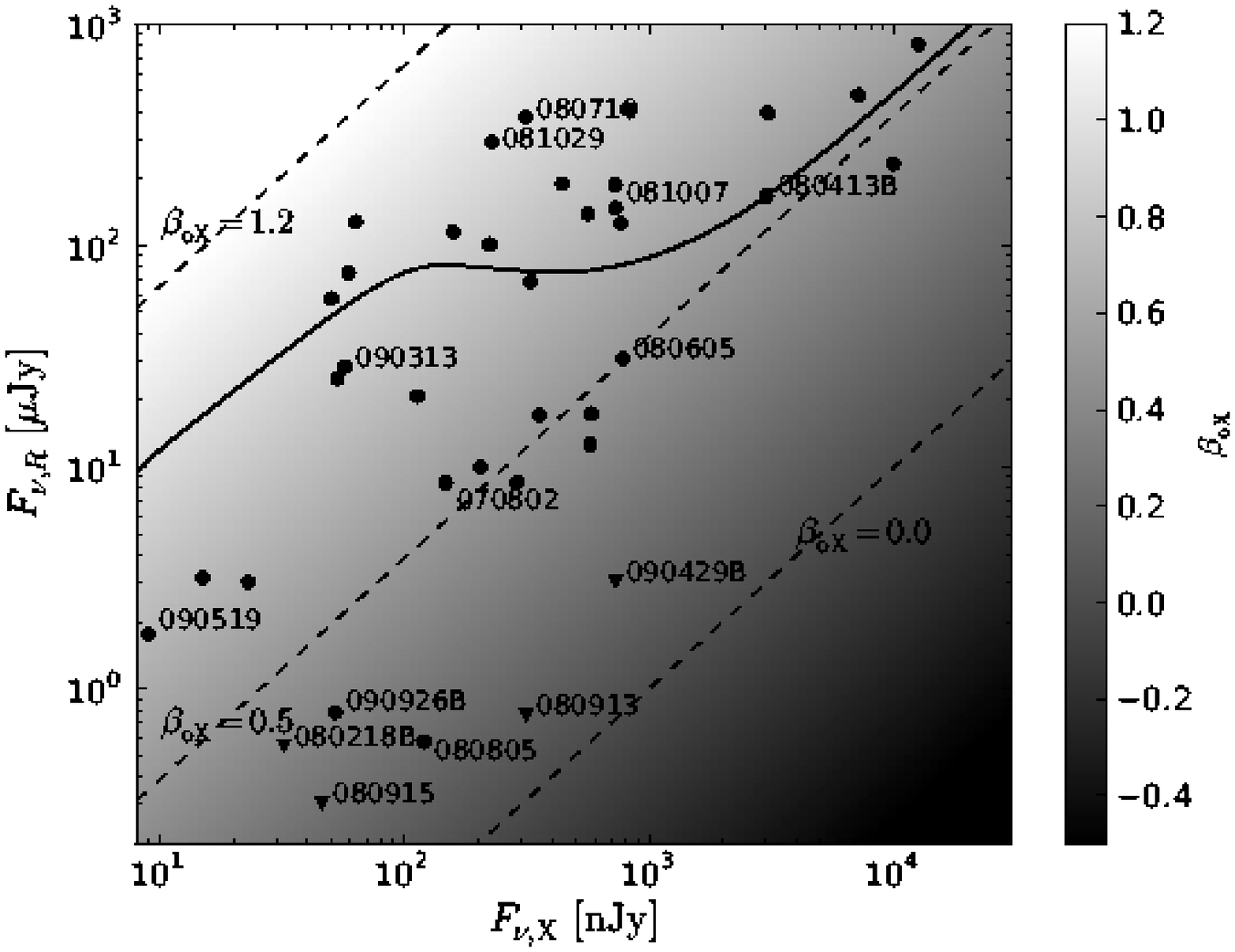}
\caption[betaox]{The distribution of our sample GRBs in the plane of $R$-band 
  flux over X-ray flux \cite{jak04}. For GRB 080413B we show as a thick
  solid line the temporal evolution of $\beta_{\rm OX}$ with time
  during the first 2 days of the afterglow evolution (from top to bottom left).
  \label{betaox}}
\end{figure}

In order to determine which of our bursts would have been classified 
as dark bursts according to the \citet{jak04} definition, we 
also estimate $\beta_{\rm OX}$ (Fig. \ref{betaox})
using the measured $R$-band brightness/limit and the X-ray flux at 3 keV
(both observer frame).
We find a fraction of 26\% 
(the nine GRBs 080218B, 080516, 080805, 080913, 080915, 090102, 090429B,
 090812, 090926B)
that would be considered "dark".
Another five bursts (070802, 080129, 080413B, 080605, 090814) have, 
within errors, all  $\beta_{OX} = 0.5$. 
According to the classification 
of \cite{hkg09}, the following nine bursts are ''dark'': 
070802, 080210, 080218B, 080516, 080805, 080913, 080915, 090429B and 090904B.
These fractions of 25--40\%  are
fully consistent with the hitherto known fraction of dark bursts
of 25\%--42\% \cite[e.g.][]{fjp09}.

There is one other problem with all of the ``dark burst'' definitions,
namely, the temporal evolution. Usually, a certain time interval is 
selected, based on the availability of optical and X-ray measurements,
and one of the ``darkness'' tests applied. However, given the increasing
sample of bursts with not just good X-ray but also good, simultaneous 
optical coverage, the complexity of light curves as compared to the
basic fireball scenario becomes more and more evident. We have seen all
kinds of different light curve behaviour, with either the optical or the
X-rays decaying faster; sometimes this even changes in one burst. As a 
consequence, there is nothing like a single $\beta_{\rm OX}$ per burst, 
but instead each burst shows a more or less pronounced evolution of 
$\beta_{\rm OX}$. As an example, the behaviour of GRB 080413B is shown in 
Fig. \ref{betaox}.

\begin{figure}[ht]
%  \vspace{-0.1cm}
  \includegraphics[angle=270,width=\columnwidth]{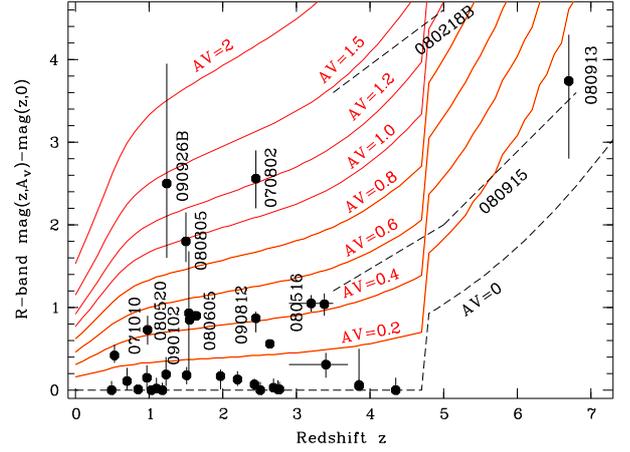}
  \vspace*{-0.1cm}
  \caption[Avredsh]{The effect of the combination of various values of 
   intrinsic extinction $A_{\rm V}$ (lines with labels) and redshift, producing
   an effective dimming in the $R$ band as given on the y-axis. 
  The solid lines have been computed assuming a dust extinction curve
   as described in \cite{Reichart1999} but with no 2175~\AA\ bump and an 
   opacity due to intergalactic hydrogen approximated by 
  $\tau ({\rm HI}) = 2.6\,\times\, (1+z)^{3.3}$ for 700 nm/$(1+z) \le 121.6$ nm
   (\citealt{Valageas1999}).
   Dots represent
   the GRBs of our sample for which we have an $A_{\rm V}$ measurement from 
   the GROND SED. 
   GRBs with effective R-band reduction of $>$0.5 mag are labeled;
   GRBs 080218B and 080915 lie above the 
   correspondingly labeled dashed lines which correspond to the pairs of
   ($z,A_{\rm V}$) values described in the text.
   Moderate $A_{\rm V}$ at moderate redshifts can easily produce
   a dimming of 1--3 mag in the $R$ band.
  \label{Avz}}
%  \vspace*{-1.cm}
\end{figure}

The main results of the SED fitting with respect to the ``dark'' burst isssue
can be summarized as follows (the following properties are non-exclusive,
and we use the definition of \cite{hkg09} in the following): 
(1) Four out of the nine dark bursts  are faint due to non-zero, but moderate
    ($A_{\rm V} \approx 0.2-1.5$), 
   extinction: 070802, 080210, 080516, 080805.
    The measured extinction in many cases appears enhanced to
   the observer due to a moderate redshift of the burst 
   (see Fig. \ref{Avz}). 
(2) One burst (090904B) is behind an additional $A_{\rm V} \sim 2$ mag dust 
   on top of the nominal $A_{\rm V} \sim 5$ Galactic foreground
   which could either be due to patchiness of the foreground or
   due to host galaxy extinction.
(3) Two bursts are faint (in $r'$) due to high redshift ($z>5$), namely 
   GRB 080913 and 090429B. 
   This corresponds to a fraction of 22\%$\pm$8\% of the dark bursts.
(4) The remaining two of our ''dark'' GRBs, 080218B and 080915, are 
   either at large redshift
   or moderate $A_{\rm V}$, or both (see notes above) -- so they belong to
   one of the above two groups (1) or (3). 
(5) Even bursts with good evidence for a spectral break and 
   $\beta_{\rm OX}$ = 0.5 do require extinction in some cases, in particular
   if the break is near the optical (rather than X-ray) range.

\begin{figure}[ht]
\includegraphics[angle=0,width=\columnwidth]{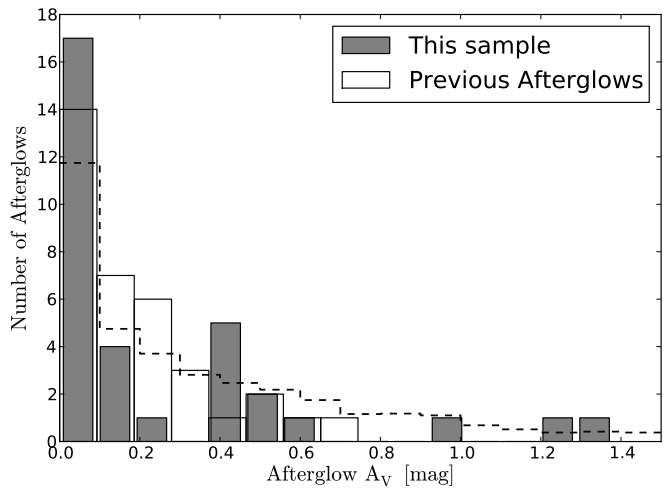}
\vspace*{-0.2cm}
\caption[Avdistr]{The distribution of measured $A_{\rm V}$ from our sample
with redshift as compared to that of \cite{kkz10}.
The dashed line is the theoretical distribution, normalized to the 
same number of objects in the sample, of a Monte Carlo simulation
of random sightlines through an evolving galaxy model \citep{uhg09}.
A KS-test returns a rejection probability of 54\% for the null
hypothesis that the two distributions are drawn from the same sample.
 \label{Avdis}}
\end{figure}

\subsection{Extinction}

The rest-frame A$_{\rm V}$ could be derived for 33 bursts of our sample
(Tab. \ref{grb30min}), and its distribution is compared against that
of the \cite{kkz10} collection (Fig. \ref{Avdis}), 
made up of all optically bright GRBs with 
photometry available from the literature. 
We find substantially larger extinction values, in particular 
about twice as many bursts ($\sim$25\% vs. 12\%)
with $A_{\rm V} \sim 0.5$ mag, and  
for the first time a significant fraction (10\%) of bursts with 
$A_{\rm V}$ \gax\ 1 mag from a direct optical/NIR afterglow
SED reddening measurement.
The difference is easily 
understood as a selection effect (predominance of bright bursts) of the 
\cite{kkz10} sample. Note, however, that this result is not due to 
the way in which the extinction models are fit to the data - our result
is derived with the canonical SMC, LMC or MW extinction laws, in the same
way as was done in \cite{kkz10}. We note in passing that using a different
fitting approach (a 'Drude' model) was reported to yield
a factor 2--5 larger visual extinction \citep{lla10}.

It is also interesting to note that the bursts with the largest
measured $A_V$ are those which tend to prefer
extinction laws that include a 2175\AA\ feature.
Unfortunately, statistics is poor, so
one has to await for future, larger samples to verify whether or not this 
is a generic trend.

Despite this larger fraction of moderate to large extinction, there is
still a 50\% fraction of bursts with $A_{\rm V} = 0$, with errors of
only $\pm$0.02--0.05 mag (Tab. \ref{grb30min}). This has also been somewhat
surprising for two reasons. First, in the canonical picture of a 
massive progenitor with a strong
wind prior to explosion, at least some extinction from the ejected
material within the wind of the progenitor is expected 
\citep[e.g.][and references therein]{wat10}.
Second, even if the local environment of the GRB is relatively dust free, 
the GRB radiation has to pass through the host galaxy to reach us.
Already the latter effect
alone should create a mean extinction of order $A_{\rm V} = 0.05-0.1$ mag 
for about 30\% of sources \citep{uhg09}, 
and very few with high A$_{\rm V}$; see Fig. \ref{Avdis}. 
We now observe both, a larger fraction of $A_{\rm V} = 0$ objects as well as
a larger fraction of $A_{\rm V} > 1$ objects
as compared to this, admittedly simplistic, expectation.
The reason for this discrepancy is presently unknown,
but could be related to the possibility that 
the dust geometry is not homogeneous, but clumped. The relative
overabundance of $A_{\rm V} = 0$ and $A_{\rm V} > 1$ objects would
then be determined by the covering factor and the dust column through
the clumps.

It is interesting to note that the $A_{\rm V}$ distribution of type II-P
supernovae \citep{sec09} (which comprises about 60\% of all core-collapse SN), 
and that of stripped-envelope supernovae \citep{rbb06} 
(which comprises 3 IIb, 11 Ib and 13 Ic) 
derived from a comparably sized sample as our GRBs, 
are consistent with being drawn from the same underlying distribution
as our $A_{\rm V}$ distribution of GRBs
(though many of their $A_{\rm V}$ values have errors consistent with
$A_{\rm V}$ = 0 at the 1$\sigma$ level).
This suggests that we sample similar environments
for both populations, hinting at similar progenitor systems: all these
supernovae tend to be associated with star-forming regions, so they can also
be expected to be significantly extinguished in their host galaxies. 
There is similar  evidence that the distribution of Wolf-Rayet
stars is consistent with the theoretical picture that type-Ic SN 
result from progenitors that have been stripped of a larger part of their
envelope \citep{lsl10}.

\subsection{Gas-to-dust ratio}

\begin{figure}[ht]
\includegraphics[angle=0,width=\columnwidth]{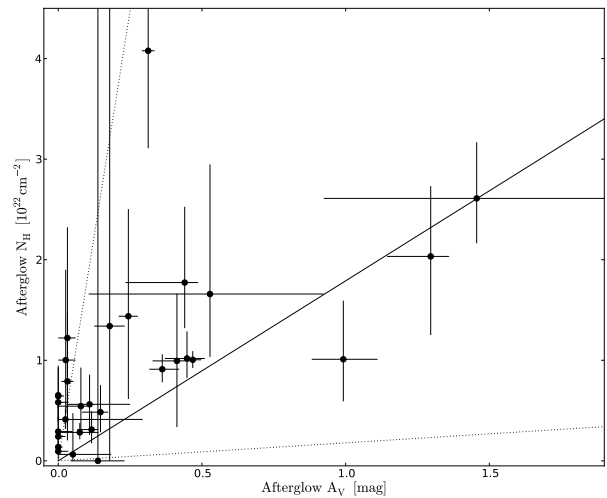}
\caption[avnh]{Relation, or lack thereof, between best-fit rest-frame 
  extinction and rest-frame neutral hydrogen absorption of our sample.
  The lines are the relation for our Galaxy \citep{ps95}, as well as
  10$\times$ and 100$\times$ larger $N_{\rm H}$ (from bottom to top).
  GRBs 080129 and 080516 are missing as they fall outside the plot with
  their large $N_{\rm H}$ values.
  \label{avnh}}
\end{figure}

Effective hydrogen absorption in excess of the Galactic foreground
absorption has taken a long time to be detected significantly in GRB 
afterglow spectra. Originally not detected at all in the full sample
of BeppoSAX bursts \citep{ppp03}, a re-analysis of the brightest
13 X-ray afterglows  revealed statistically significant absorption in excess 
of the Galactic one for two bursts \citep{sfa04}. Already 8 bursts of 17
observed with {\it Chandra} or XMM-{\it Newton} up to Oct. 2004 show excess 
absorption \citep{gcp06}. In a systematic study of 93 promptly observed
{\it Swift} GRBs  with known redshift (up to May 2009), 85 show evidence
of intrinsic X-ray absorption at the host galaxy site \citep{ctu10}.
Similarly, in our sample, we detect excess absorption in 26 
out of 33 cases. This difference in the excess absorption detection rate is 
primarily related to the quality of the X-ray spectra, which pre-{\it Swift} 
was typically taken 8--12 hours post trigger, by which time the signal-to-noise
ratio was insufficiently high to accurately measure any intrinsic absorption.

Figure \ref{avnh} shows a comparison of the effective neutral hydrogen 
absorption $N_{\rm H}$ and visual extinction $A_{\rm V}$, both in the 
rest frame of the corresponding burst. As has been noted frequently in
the past \citep[e.g.][]{gaw01, sfa04, spo10}, 
the $N_{\rm H}$-to-$A_{\rm V}$ ratio is far from being similar
among different bursts, and substantially larger than in our Galaxy.
Note, however, that here, as has been usual in previous cases, solar 
metallicity has been assumed in deriving $N_{\rm H}$. Since the observed 
curvature in the X-ray spectra is predominantly, but not exclusively, due to
absorption by oxygen, the derived effective $N_{\rm H}$ is inversely
proportional to the metallicity (or better O/H ratio) of the burst 
environment. Since this
metallicity has been observed (in other GRBs) to be about 1/5
solar (though extremes of 
solar \citep{psp09,efh09} up to super-solar \citep{srg10} and nearly 
1/100 solar \citep{dfm07, rsk10} do also occur), the effective $N_{\rm H}$
would likely be even larger than shown in Fig. \ref{avnh}, if the proper
line-of-sight metallicity were to be used (if it were known).

In contrast to our $A_{\rm V}$ distribution,
the distribution of $N_{\rm H,X}$ from the complete {\it Swift} sample lacks
a substantial fraction of zero column density \citep{ctu10}. This has
been explained by \cite{ctu10} as evidence that the bursts originate
within high-density regions of their hosts, since a random distribution
in a galaxy like ours would predict a sizable fraction ($\sim$30\%)
with no intrinsic absorption. By combining their sample
with the Lyman-$\alpha$ absorbers at $z>2$ of \cite{fjp09} they also find, 
similar to earlier reports \citep[e.g.][]{whf07}, 
that the bulk of GRBs have column densities in X-rays which are
a factor $\sim$10 higher than in the optical ($N_{\rm HI}$), which they 
explain by ionization of hydrogen by the high energy flux of the GRB.
Since this ratio is roughly similar to that of $N_{\rm H,X}$ vs. $A_V$ 
(Fig. \ref{avnh}), one could think that $A_V$ and the 
Lyman-$\alpha$ absorption are correlated. However, the three bursts with
reported $N_{\rm HI}$ in \cite{fjp09} (070802, 071031 and 080804)
do not show any correlation, similar to the 6 bursts from the UVOT
sample published by \cite{spo10}.

\subsection{Redshift distribution}

\begin{figure}[ht]
\includegraphics[angle=270,width=\columnwidth]{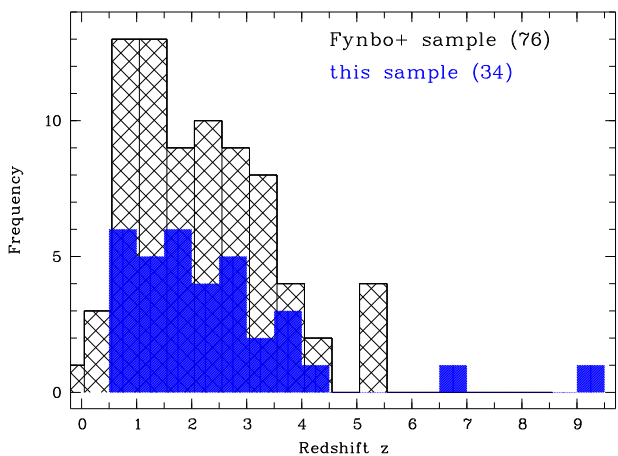}
\caption[zdis]{Redshift distribution of our GROND sub-sample (blue) vs. that of
the hitherto most complete sample of \citep{fjp09} (bursts with
upper limits have been omitted).
  \label{zdis}}
\end{figure}

The redshift distribution of our sample is 92\% complete
(also one of the 4 GRBs not detected by GROND has a redshift).
We re-iterate that the only selection criterion was the detection of
an X-ray afterglow, and do not see a bias introduced by the requirement
of a rapid GROND observation (or equivalently an occurrence during
Chilean night time).
A comparison to the distribution of all known long-duration bursts 
(about 50\% complete; Fig. \ref{zdis}) reveals the former to have a flatter 
distribution, with a somewhat higher number of $z>4$ bursts. 
However, a KS-test shows that this is not statistically significant,
and both distributions are consistent
with being drawn from the same sample within 1$\sigma$.

The presence of GRBs 080913 and 090429B in our sample corresponds to 
a fraction of 5.5$\pm$2.8\% of bursts at redshifts $z>5$, 
and the strict upper limit would be 12.8\% if all three GRBs 
without redshift would be at $z>5$. 
A larger sample size with a similarly good completeness level would
be required to derive fractions with errors less than the present 
$\sim$50\% level.

\section{Conclusions}

The majority of afterglow SEDs show a spectral break between the X-ray and
optical/NIR range 
that can be described with a slope difference of 0.5, consistent with the
basic fireball scenario. This spectral break implies a $R$-band flux of 
about 3--4 mag fainter than obtained from an extrapolation of the 
X-ray spectrum. This effect is dealt with in all the recent definitions
of ``darkness'' \citep[e.g.][]{phh06}, 
and our finding of a dominance of SED breaks is consistent 
with the large number of bursts seen at $\beta_{\rm O} < 1$.

The faint optical afterglow emission of ``dark bursts'', where we used
the definition of \cite{hkg09},  is due to a mixture of
moderate intrinsic extinction at moderate 
redshifts, and a fraction of bursts at redshift $>$5 (about 25\% of the dark
bursts).

This finding is in line with previous investigations 
\citep{mmk08, ckh09, pcb09, fjp09}.
In particular, \cite{ckh09} used a similar approach as ours,
namely a sample of 29 bursts for which follow-up observations
with the robotic Palomar 60\,inch telescope began within 1 hr after the
burst trigger. They recovered 80\% of the optical afterglows, as compared
to our 90\% with a $\sim$60\% fraction of redshifts (as compared to our 92\%).
A search for host galaxies was then performed for bursts without afterglow
and/or redshift of this sample to assess the degeneracy
between high-$z$/low-$A_{\rm V}$ vs. low-$z$/high-$A_{\rm V}$ \citep{pcb09}.
In combination, and assuming further that the extinction measured along
the line-of-sight to GRBs is proportional to the extinction of the
host emission, these authors
reach the similar conclusion that extinction
is responsible in large part for the ``dark bursts''.
Some of their $A_V$ values, however, have been derived by assuming a
SED slope in the optical, rather than measuring it. Given the prevalence
of breaks (see above), this is risky.

In contrast, we emphasize here, that we (1) measure the optical/NIR SED,
and (2) make no assumptions on the host properties. Thus, our sample
of bursts is the first with properly measured $A_{\rm V}$ values
which neither requires a relative shifting of different filter 
measurements, as often done in the past nor suffers from the small wavelength
coverage. Our only assumption is that whenever
the data prefer a break in the SED between X-rays and optical/NIR, we
only allow a slope difference of 0.5. GROND observations start removing
the bias of previous studies to bright bursts, i.e. low-$A_{\rm V}$ and 
low-$z$, and start to detect and reliably measure these higher-$A_{\rm V}$
values directly. From individual bursts not in this sample there is
evidence that bursts with even  $A_{\rm V} \sim 3-5$ exist
(e.g. GRB 080607, \cite{psp09}; GRB 070306, \cite{jrw08}) 
-- though these are admittedly rare
occasions which require larger samples to include them.

\begin{acknowledgement}
We thank S. Savaglio for useful comments. TK acknowledges support by
the DFG cluster of excellence 'Origin and Structure of the Universe',
and AU is grateful for travel funding support through MPE.
FOE acknowledges funding of his Ph.D. through the 
\emph{Deutscher Akademischer Austausch-Dienst} (DAAD),
SK and ARossi acknowledge support by DFG grant Kl 766/13-2 
and ARossi
additionally from the BLANCEFLOR Boncompagni-Ludovisi, n\'ee Bildt foundation.
MN acknowledges support by DFG grant SA 2001/2-1.
Part of the funding for GROND (both hardware as well as personnel)
was generously granted from the Leibniz-Prize to Prof. G. Hasinger
(DFG grant HA 1850/28-1).
This work made use of data supplied by the UK {\it Swift} Science Data Centre 
at the University of Leicester.
\end{acknowledgement}

\bigskip

%{\it Facilities:} \facility{Max Planck:2.2m (GROND)}, 
%                  \facility{Gemini:South (GMOS)},
%                  \facility{SAAO:Nagoya (SIRIUS)},
%                  \facility{INTEGRAL (SPI-ACS)}

\noindent {\small {\it Facilities:} Max Planck:2.2m (GROND), 
                  Swift}

\end{document}